\documentclass[aps,twocolumn,preprintnumbers,superscriptaddress]{revtex4-1}
\usepackage[T1]{fontenc} % Use Type1 fonts
\usepackage[utf8]{inputenc}
\usepackage{multirow}
\usepackage{float}
\usepackage{bm}% bold math
\usepackage{booktabs} % Use booktabs and array for better tables
\usepackage{array} %q
\usepackage{graphicx}
\usepackage{amsmath}
\usepackage{amssymb}
\usepackage{color}
\usepackage[per-mode=symbol,separate-uncertainty]{siunitx} %Units package.
\usepackage[capitalize]{cleveref} %Better references.
\usepackage{physics} % For absolute values and norms 
\usepackage{array,multirow,makecell}
\usepackage[table]{xcolor}

\setcellgapes{1pt}
\makegapedcells
\newcolumntype{R}[1]{>{\raggedleft\arraybackslash }b{#1}}
\newcolumntype{L}[1]{>{\raggedright\arraybackslash }b{#1}}
\newcolumntype{C}[1]{>{\centering\arraybackslash }b{#1}}

\newcount\colveccount
\newcommand*\colvec[1]{
         \global\colveccount#1
         \begin{pmatrix}
         \colvecnext
}
\def\colvecnext#1{
         #1
         \global\advance\colveccount-1
         \ifnum\colveccount>0
                 \\
                 \expandafter\colvecnext
         \else
                 \end{pmatrix}
         \fi
}

\begin{document}

\author{Luca Planat}
\author{R\'emy Dassonneville}
\author{Javier Puertas Mart\'inez}
\author{Farshad Foroughi} 
\author{Olivier Buisson} 
\author{Wiebke Hasch-Guichard} 
\author{C\'ecile Naud}
\affiliation{Univ. Grenoble Alpes, CNRS, Grenoble INP, Institut N\'eel, 38000 Grenoble, France}
\author{R. Vijay} 
\affiliation{Tata Institute of Fundamental Research, Mumbai, India}
\author{Kater Murch}
\affiliation{Washington University in St. Louis, USA}
\author{Nicolas Roch}
\affiliation{Univ. Grenoble Alpes, CNRS, Grenoble INP, Institut N\'eel, 38000 Grenoble, France}
\email{nicolas.roch@neel.cnrs.fr}

\title{Understanding the saturation power of Josephson Parametric Amplifiers made from SQUIDs arrays}

\begin{abstract}
We report on the implementation and detailed modelling of a Josephson Parametric Amplifier (JPA) made from an array of eighty Superconducting QUantum Interference Devices (SQUIDs), forming a non-linear quarter-wave resonator. This device was fabricated using a very simple single step fabrication process. It shows a large bandwidth (\SI{45}{\mega\hertz}), an operating frequency tunable between \SI{5.9}{\giga\hertz} and \SI{6.8}{\giga\hertz} and a large input saturation power ($-117\ \text{dBm}$) when biased to obtain \SI{20}{\decibel} of gain. Despite the length of the SQUID array being comparable to the wavelength, we present a model based on an effective non-linear LC series resonator that quantitatively describes these figures of merit without fitting parameters. Our work illustrates the advantage of using array-based JPA since a single-SQUID device showing the same bandwidth and resonant frequency would display a saturation power \SI{15}{dB} lower. \end{abstract}

\maketitle

\section{Introduction}

Gain, bandwidth, and noise performance ultimately dictate the quantum efficiency and  speed of quantum measurements performed at microwave frequencies in the circuit quantum electrodynamics architecture. Improving these three properties of an amplifier has been the driving force for Josephson junction based amplifier design and characterisation including significant work optimising bandwidth~\cite{Mutus:2014dd, Roy:2015ky}, pump rejection using flux or non-degenerate pumping schemes~\cite{Yamamoto:2008cr, Bergeal:2010iu, Roch:2012gy, Mutus:2013iw, Eichler:2014dk, Frattini:2017ji} and realizing directionality in the amplification process~\cite{Abdo:2013ib,Sliwa:2015vc,Lecocq:2017ge}. Josephson junction based parametric amplifiers utilise the intrinsic nonlinearity of the junction as the basis for parametric wave mixing. Controlling the type and strength of this nonlinearity has been the focus of several amplifier designs since this quantity imposes the input saturation power (characterised as the $\SI{1}{\decibel}$ compression point) of such amplifiers~\cite{Abdo:2011dfa,Eichler:2013cr,Eichler:2014iw,Zhou:2014gt,Eddins:2017ty,Liu:2017bi,Frattini:2018ud}. Moreover, when the strength of the non-linearity reaches a few percent of the operating frequency of the device, higher-order effects lead to imperfect squeezing and non quantum-limited performance~\cite{Yurke:2006cw,Kochetov:2015tp,Boutin:2017dn}. However it is only very recently that the fourth order non-linearity or Kerr non-linearity was identified as the main cause of Josephson parametric amplifiers saturation~\cite{Liu:2017bi}. In their work, \textit{Liu et  al.} didn't manage to relate the effective non-linearity of their Josephson junction amplifier to the actual circuit model. This outstanding goal was achieved soon after in the case of the SNAIL Parametric Amplifier, a Josephson device operated in a three-wave mixing mode~\cite{Frattini:2018ud}. In this article we present a parametric amplifier based on four-wave mixing. The subtlety here is that the non-linearity at the root of parametric amplification is the same than the one causing saturation. Our device is formed by a high impedance Josephson meta-material --- an array of $N=80$ SQUIDs ---  that forms a $\lambda/4$ non-linear resonator~\cite{YURKE:1996dt, CastellanosBeltran:2008cg, Anonymous:HxLrydHU, Vesterinen:2017ei}. The dispersion relation of this SQUIDs array, obtained via two tone spectroscopy, is fitted using a long range Coulomb interaction or \textit{remote ground} model~\cite{krupko2018kerr} leading to independently inferred values of the circuit components (capacitances and SQUIDs' critical current). We show that the amplifier can be quantitatively described by an effective non-linear $LC$ series resonator with a resonant frequency near the first resonant mode of the array. We report a good quantitative agreement between the saturation power of this JPA and a model without fitting parameters. According to this theory, the 80 SQUID array yields a \SI{15}{\decibel} improvement over a comparable single SQUID device.

This article is organized as follow: In Section II we present an effective model and review the basic description of our device as a single port degenerate Josephson parametric amplifier and in section III we discuss how arrays of SQUIDs can effectively reduce the nonlinearity of the device leading to increased saturation power.  In Section IV we present the device and Section V describes its properties in the linear regime. Gain and saturation are reported in Section VI while we discuss the main results in section VII.

\section{Model}

\begin{figure}[h]
\includegraphics[width=\linewidth]{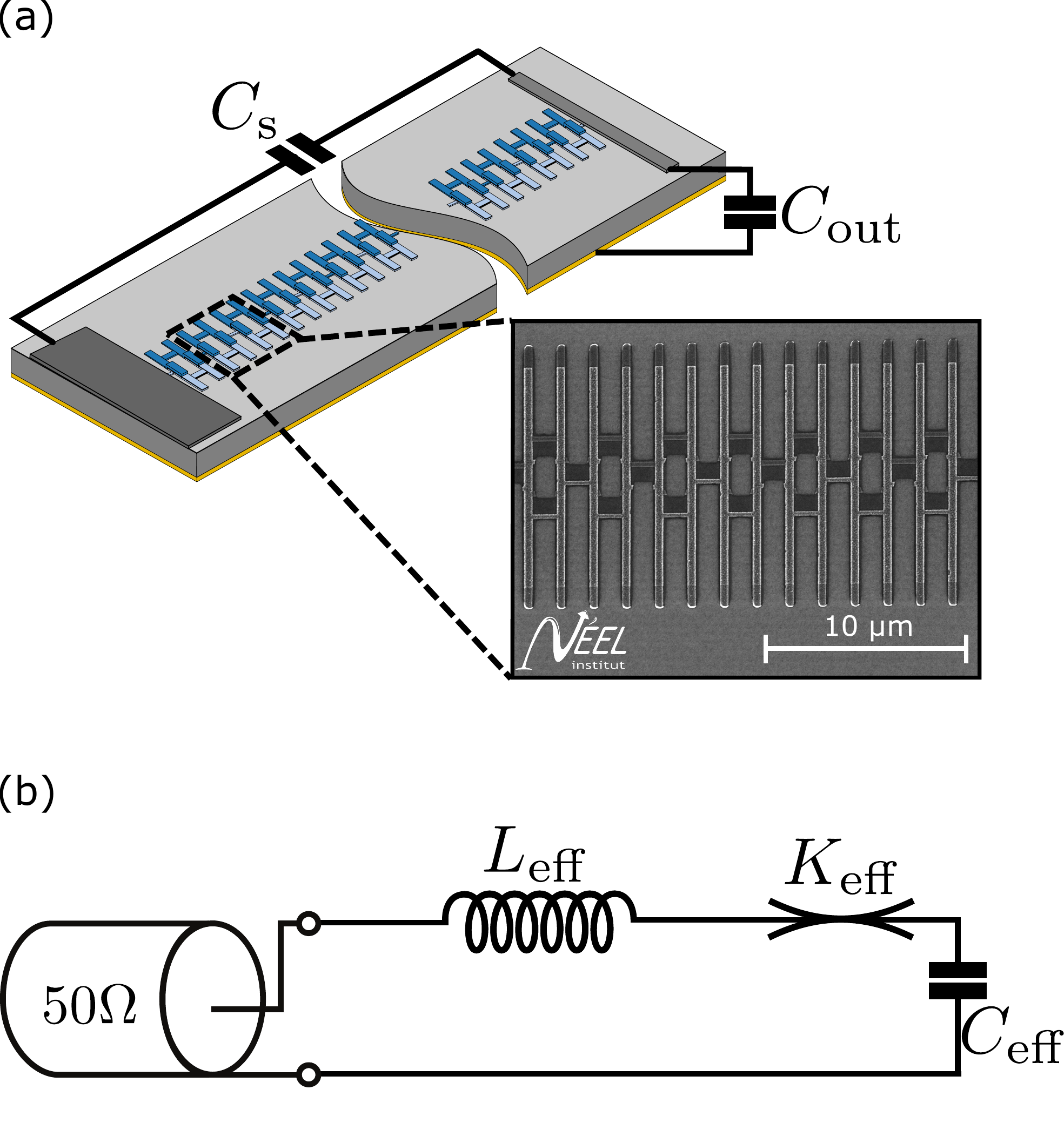}
\caption{\textbf{a} Sketch of the JPA based on an array of 80 SQUIDs. We highlight the capacitance $C_\text{out}$ between the last superconducting pad and the ground and the parasitic shunt capacitance  $C_\text{s}$ between the input/output pad and the last superconducting pad. (Zoom-in) SEM image of 7 identical SQUIDs where a single junction has an area of $\SI{10.7}{\micro\meter} \times \SI{0.370}{\micro\meter}$. \textbf{b} Effective LC series nonlinear resonator.}
\label{fig2}
\end{figure}
\begin{figure}[h]
\includegraphics[width=\linewidth]{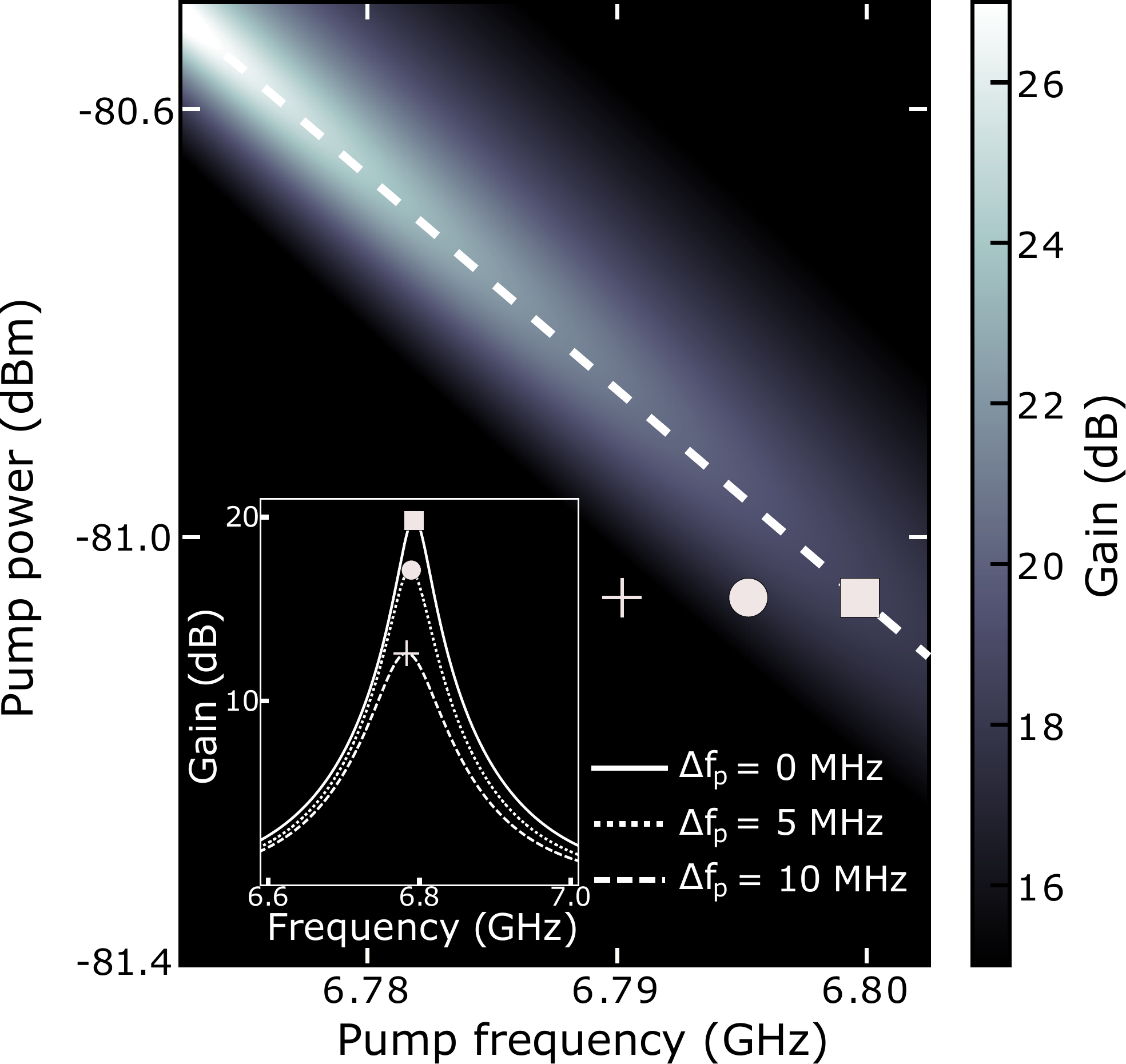}
\caption{
Theoretical maximum signal gain $|g_{S,\Delta}|^2$ versus pump power and pump frequency for signal detuning $\Delta=0$. Larger gain requires larger pump power and lower pump frequency. The inset shows the gain versus frequency for three different pump configurations as indicated by the markers. The maximum expected gain for a given pump power (white square) is reduced when the pump frequency is slightly detuned from this optimal point (circle and cross). }
\label{fig1}
\end{figure}

Our device (See \cref{fig2}) can be modeled as an effective single port, degenerate Josephson parametric amplifier employing a non-linearity of the Kerr type. The circuit can be described by the Hamiltonian of a non-linear resonator:
\begin{equation}
H_\text{JPA}=\hbar\omega_\text{eff}A^\dagger A+\hbar\frac{K_\text{eff}}{2}\left (A^\dagger\right)^2 A^2
\end{equation}
where $A$ is the annihilation operator of the intra-resonator field. It is characterised by an effective resonant frequency $\omega_\text{eff}$, a non-linearity or self-Kerr coefficient $K_\text{eff}$ and a coupling rate to a transmission line $\kappa_\text{eff}$. The link between the circuit model and these effective parameters will be explained in Section V. The JPA is powered by a monochromatic current pump. The physics of such degenerate Josephson parametric amplifiers has been explained in great detail in various articles~\cite{Yurke:2006cw,Vijay:2009wp, Roy:2015ky, Boutin:2017dn}. We recall here the main equations, following the approach of \textit{Eichler and Wallraff}~\cite{Eichler:2014iw}.  The dynamics of the circuit is inferred using conventional input-output theory:
\begin{equation}
\dot{A} = -i\omega_\text{eff}A-iK_\text{eff}A^\dagger AA - \frac{\kappa_\text{eff}}{2}A+\sqrt{\kappa_\text{eff}} A_\text{in}
\label{EOM}
\end{equation}
with $A_\text{in}$ the input field coupled with rate $\kappa_\text{eff}$. The boundary conditions of the resonator are taken into account via the equation $A_\text{out} = \sqrt{\kappa_\text{eff}}A - A_\text{in}$, where $A_\text{out}$ is the output field. Next we assume that $A=\alpha+\hat{a}$, where $\alpha$ is a classical part (referring to the strong coherent pump) and $\hat{a}$ is the signal that we treat quantum mechanically. To obtain the gain of the amplifier we follow a two-step procedure: we first solve for the classical field $\alpha$ while setting $\langle\hat{a}\rangle=0$ and then we use a linearisation of the equation of motion for $\hat{a}$ around this working point (see Appendix~\ref{sec:gain} for a detailed derivation). This leads to the standard equation of a parametric amplifier:
\begin{equation}
\hat{a}_\text{out}(\Delta)=g_{S,\Delta}\hat{a}_\text{in}(\Delta)+g_{I,\Delta}\hat{a}_\text{in}^\dagger(-\Delta)
\label{gain}
\end{equation}
The operators $\hat{a}_\text{in}(\Delta)$ and $\hat{a}_\text{out}(\Delta)$ are the Fourier components of the input and outputs signals, where $\Delta$ is the dimensionless frequency detuning $\Delta = (\omega_\text{p} - \omega_\text{signal})/\kappa$ from the pump frequency. \cref{gain} illustrates the link between the output signal and the inputs at signal ($\Delta$) and idler $(-\Delta)$ frequencies. Signal and idler gain ($g_{S,\Delta}$ and $g_{I,\Delta}$, respectively) are expressed as:
\begin{subequations}
\begin{align}
g_{S,\Delta} &= -1 + \frac{i(\delta - 2\xi_\text{$\alpha$} n - \Delta) + \frac{1}{2}}{(i\Delta - \lambda_{-})(i\Delta - \lambda_{+})} \\
g_{I,\Delta} &= \frac{-i\xi_\text{$\alpha$} n e^{2i\phi}}{(i\Delta - \lambda_{-})(i\Delta - \lambda_{+})}  
\end{align}
\label{gsgi}
\end{subequations}
with $\lambda_{\pm}=\frac{1}{2} \pm \sqrt{((\xi_\text{$\alpha$} n)^2 - (\delta - 2\xi_\text{$\alpha$}n))}$ (a complete derivation is given in Appendix~\ref{sec:gain}). Both of them encompass the dimensionless detuning $\delta$ between the pump and bare resonator frequencies, $\xi_\text{$\alpha$}$ the product between the non-linearity and the pump power, $n$ the normalised number of pump photons in the cavity, $\phi$ the phase difference between the pump and the signal and $\Delta$. The exact expressions for these parameters are given in Appendix~\ref{sec:gain}. $|g_{S,\Delta}|^2$ is plotted in \cref{fig1}, using the parameters of our amplifier, as a function of the pump power and pump frequency for zero detuning between the pump and the signal ($\Delta = 0$). We define the optimal pump frequency $f_\text{p,opt}$ as the one which maximises the gain for a given pump power, as shown in \cref{gsgi}.
$\Delta f_p=f_\text{p}-f_\text{p,opt}$ is a frequency shift from optimal pumping conditions. As illustrated on the inset of \cref{fig1}, $\Delta f_p$ as small as $\SI{5}{\mega\hertz}$ leads to a reduction of the gain in excess of $\SI{1}{\decibel}$. This observation is at the heart of JPA saturation~\cite{Liu:2017bi}, since an input power of $n_{s}$ signal photons per second will lead to a shift $\Delta f_p \approx n_{s} \times K_\text{eff}/\kappa_\text{eff}$ Hertz from optimal pumping frequency, thus leading to a drop of the maximum expected gain. This qualitative explanation will be further formalised in Section VI and it can be shown that the input saturation power of JPA increases linearly with the ratio $\kappa_\text{eff}/\abs{K_\text{eff}}$~\cite{Eichler:2014iw}. Maximizing this ratio, and thus minimizing the non-linear self-Kerr term $K_\text{eff}$, is therefore of prime importance.

\section{Decreasing non-linearity using arrays}\label{nonlinearity}

Because reduced $K_\text{eff}$ is important to improve dynamical range, several methods have been introduced to reduce non-linearity in parametric amplifiers. One first option is to use the intrinsic non-linearity of superconductors such as Niobium~\cite{Tholen:2007ea}, NbN~\cite{Chin:ws} or granular Aluminium~\cite{Maleeva:2018uya}, since they often come with very weak non-linearities spanning from $\SI{20}{\milli\hertz}$ to $\SI{30}{\kilo\hertz}$ for resonators in the $\SI{}{\giga\hertz}$ range. However, these non-linearities are so weak that extremely large pump powers are required resulting in experimental challenges. Another option is to dilute the non-linearity of a single Josephson junction into a larger and linear resonator~\cite{Bourassa:2012ej,Zhou:2014gt}. However, in this case the Josephson junction is not purely phase-biased anymore and the usual quartic approximation used to treat the Josephson potential has a limited validity~\cite{Eichler:2014iw,Boutin:2017dn}.
Already in the early days of Josephson Parametric Amplifiers it was recognized that using an array of $N$ Josephson junctions could be beneficial~\cite{YURKE:1996dt}. In this case the total phase drop $\Phi_{tot}$ (or equivalently the voltage drop) occurs across the whole array and not across a single junction. Thus the non-linearity is divided by $N$ since each junction experiences a phase drop $\Phi_{tot}/N$ (See Appendix~\ref{sec:nonlin} for a derivation). This idea can be pushed further by fabricating an array of $N$ Josephson junctions with critical current $N$ times larger; the non-linearity is then divided by $N^2$~\cite{Eichler:2012ixa,Eichler:2014iw,Zhou:2014gt}. However, the approximation that each junction experiences a $\Phi_{tot}/N$ phase drop loses validity when the system becomes very large, reaching a size comparable to the wavelength of the microwave signal. In this case propagating effects should be accounted for properly. 

To do so we start by defining the normal modes of the circuit and then treat the non-linearity perturbatively as described in previous works~\cite{Anonymous:2012ek,Weissl:2015do, Roy:2017wp}. Each SQUID is considered as an $LC$ parallel oscillator, described by $C_\text{J}$ and $L_\text{J}$. However, describing the chain as a standard transmission line as it is routinely done, where every $LC$ oscillator is shunted to the ground via a ground capacitance $C_g$ is not the most appropriate description for our system. Given that the distance between the chain and the ground plane is comparable or greater than the modes wavelength (see \cref{fig2}.\textbf{a}), the screening of the charges by the ground plane cannot be considered as local. Capacitive effects between SQUIDs must be accounted for via the long-range part of the Coulomb interaction. We follow the procedure described by \textit{Krupko et al}~\cite{krupko2018kerr} to take this long-range interaction into consideration. This remote ground model gives results closer to the experiment than the standard transmission line model (see Appendix \ref{sec:Remote}). Although this remote ground model is more complex than the standard model, there is still only one parameter describing the screening effect: it is no longer the ground capacitance $C_g$ but a characteristic length of the long range Coulomb interaction preventing from a divergence of the model, called $a_0$. In the description of the capacitive effects in our amplifier, we also consider that the chain is terminated by a metallic pad creating an additional capacitance to ground $C_\text{out}$ and a shunt capacitance $C_{s}$ together with the input transmission line (see figure \cref{fig2}.\textbf{a}). More specifically the system is modelled by considering the Lagrangian $\mathcal{L}$ of the chain, where the fluxes $\Phi_{n}$ between each SQUID are taken as coordinates. This Lagrangian reads:
\begin{equation}
\begin{split}
\mathcal{L} = \sum\limits_{n=0}^{N-1}{\frac{C_J}{2}(\dot{\Phi}_{n+1}  - \dot{\Phi}_{n} )^2} -\sum\limits_{n=0}^{N-1}{\frac{1}{2L_J}(\Phi_{n+1} - \Phi_{n})^2} \\ 
 + \sum\limits_{n=1}^{N-1}{\frac{C_{g,nn}}{2}\dot{\Phi}_n^2} + \sum\limits_{n=1}^{N-1}{\sum\limits_{i\neq n}^{N-1}{\frac{C_{g,ni}}{2}(\dot{\Phi}_n^2 - \dot{\Phi}_i^2}})   \\
 + \frac{C_{out}}{2}\dot{\Phi}_N^2 + \frac{C_s}{2}\dot{\Phi}_N^2
\end{split}
\end{equation}
with $N$ the number of SQUIDs in the chain and $C_{g,ni}$ are the elements of a generalised ground capacitance matrix. We define a new set of variables to describe the system, the charge $Q_n$ and its conjugate $I_n$ at each node $n$
\begin{equation}
\begin{aligned}
Q_n = \frac{\partial{\mathcal{L}}}{\partial{\dot{\Phi}_n } } \\
I_n = \frac{\partial{\mathcal{L}}}{\partial{\Phi_n } }
\end{aligned}
\end{equation}
These new variables lead to capacitance and inductance matrices ($\hat{C}$ and $\hat{L}$ respectively): 
\begin{equation}
\begin{aligned}
\vec{Q} = \hat{C}\vec{\dot{\Phi}}  \\
\vec{I} = \hat{L}^{-1}\vec{\Phi}.
\end{aligned}
\end{equation}
From these matrices, we can define the angular frequency matrix $\hat{\Omega}$ as :
\begin{equation}
\hat{\Omega}^2=-\hat{C}^{-1}\hat{L}^{-1}
\end{equation}
The eigenvalues $\omega_{n}^2$ and eigenvectors of the matrix $\hat{\Omega}^2$ define respectively the resonant frequency and the wave profile of each mode $n$ of the chain. It allows the definition of an effective capacitance $C_{\text{eff},n}$ and an effective inductance $L_{\text{eff},n}$ for each mode $n$:
\begin{equation}
\begin{aligned}
C_{\text{eff},n} = \vec{\varphi}_{n}^T\hat{C}\vec{\varphi}_{n} \\
L_{\text{eff},n}^{-1} = \vec{\varphi}_{n}^T\hat{L}^{-1}\vec{\varphi}_{n}
\end{aligned}
\end{equation}
With this linear model, we now treat the Kerr non-linearity of the chain. The Josephson non-linearity can be reintroduced as a perturbation of the linear Hamiltonian, by developing the cosine of the Josephson potential up to fourth order~\cite{krupko2018kerr}. By applying the Rotating Wave Approximation (RWA), one can rewrite the full Hamiltonian as: 
\begin{equation}
\begin{split}
\hat{H} = \sum\limits_{n}{\hbar\omega_{n}a^{\dagger}_na_n} -\sum\limits_{n}{\frac{\hbar}{2}K_{nn}a^{\dagger}_na_na^{\dagger}_na_n}  \\
-\sum\limits_{n,m}{\frac{\hbar}{2}K_{nm}a^{\dagger}_na_na^{\dagger}_ma_m}
\end{split}
\end{equation}
where $K_{nn}$ and  $K_{nm}$ are the self and cross Kerr coefficients, respectively:
\begin{equation}
\begin{aligned}
K_{nn} = \frac{2\hbar\pi^4E_\text{J}\eta_{nnnn}}{\Phi_{0}^4C_\text{J}^2\omega_{n}^2} \\
K_{nm} = \frac{4\hbar\pi^4E_\text{J}\eta_{nnmm}}{\Phi_{0}^4C_\text{J}^2\omega_{n}\omega_{m}}
\end{aligned}
\end{equation}
$\eta_{nnmm}$ takes into account the spatial variation of the phase across the chain and $E_\text{J}=\varphi_\text{o}^2/L_\text{J}$ is the Josephson energy of a single SQUID. Given that $\eta_{nnmm}$ depends only on circuit parameters of the chain, the Kerr non-linearities of the modes are fully predictable.
To describe the effect of the transmission line connected to the array and the resulting external quality factor, we model this $\lambda/4$ resonator as an effective non-linear series $LC$ circuit (See \cref{fig2}.\textbf{b}) close to its resonance. From now on, we drop the index $n$ since we only consider the first mode. Using the effective inductance and capacitance defined previously, we can then easily define an effective resonant frequency $\omega_\text{eff}=1/\sqrt{L_\text{eff}C_\text{eff}}$, an effective external quality factor $Q_\text{eff}=\sqrt{L_\text{eff}/C_\text{eff}}/Z_\text{c}$ and an effective coupling rate $\kappa_\text{eff}=\omega_\text{eff}/Q_\text{eff}$, as is very commonly done in microwave engineering~\cite{pozar2009microwave}. The accuracy of this mapping relies on a precise determination of the capacitance and inductance matrices. Using a combination of electromagnetic simulations and two-tone measurements we managed to infer precisely $\hat{C}$ and $\hat{L}$ as will be explained later.

\section{Sample Description}\label{sample}

The JPA presented in this work is made of 80 SQUIDs, obtained using a bridge-free fabrication technique~\cite{Lecocq:2011dk}. It is fabricated on a $\SI{300}{\micro\meter}$ thick, single side polished, intrinsic silicon wafer. The backside of the wafer is metalized using a sandwich of titanium ($\SI{10}{\nano\meter}$) and gold ($\SI{200}{\nano\meter}$). The array is connected galvanically to a $\SI{50}{\ohm}$ microstrip transmission line on one side and to a superconducting pad on the other side (\cref{fig2}.\textbf{a}). Such a design presents two main advantages. It can be fabricated in one single electronic lithography step and since no superconducting ground plane is involved, flux-trapping possibilities and the effect of Meissner currents are strongly reduced.
The associated circuit parameters are $C_\text{J}=\SI{370}{\femto\farad}$, $a_0=\SI{4.3}{\micro\meter}$, $C_\text{out}=\SI{24.7}{\femto\farad}$ $C_\text{s}= \SI{1}{\femto\farad}$. Finally, $L_\text{J}=\SI{165}{\pico\henry}$ at zero magnetic flux,  which translates into a critical current of $I_\text{c}=\SI{2}{\micro\ampere}$ for each SQUID. $L_\text{J}$ remains larger than the kinetic inductance of the aluminum wires connecting the SQUIDs. We estimate this stray inductance to be $L_\text{stray}=\SI{30}{\pico\henry}$. Ensuring $L_\text{stray}\ll L_\text{J} $ is important to the validity of our model (\cref{fig2}.\textbf{b}). $C_\text{J}$ is inferred via the size of the junctions and the capacitance density $\SI{45}{\femto\farad\per\square\micro\meter}$~\cite{Fay:2008vi}. The values of $C_\text{out}$ and $C_\text{s}$ are obtained using an electromagnetic simulation software. $L_\text{J}$ and $a_0$ are determined from the dispersion relation of the array, as explained in Section~\ref{linear}.

\section{Characterization in the linear regime}\label{linear}

\begin{figure}[h]
\includegraphics[width=\linewidth]{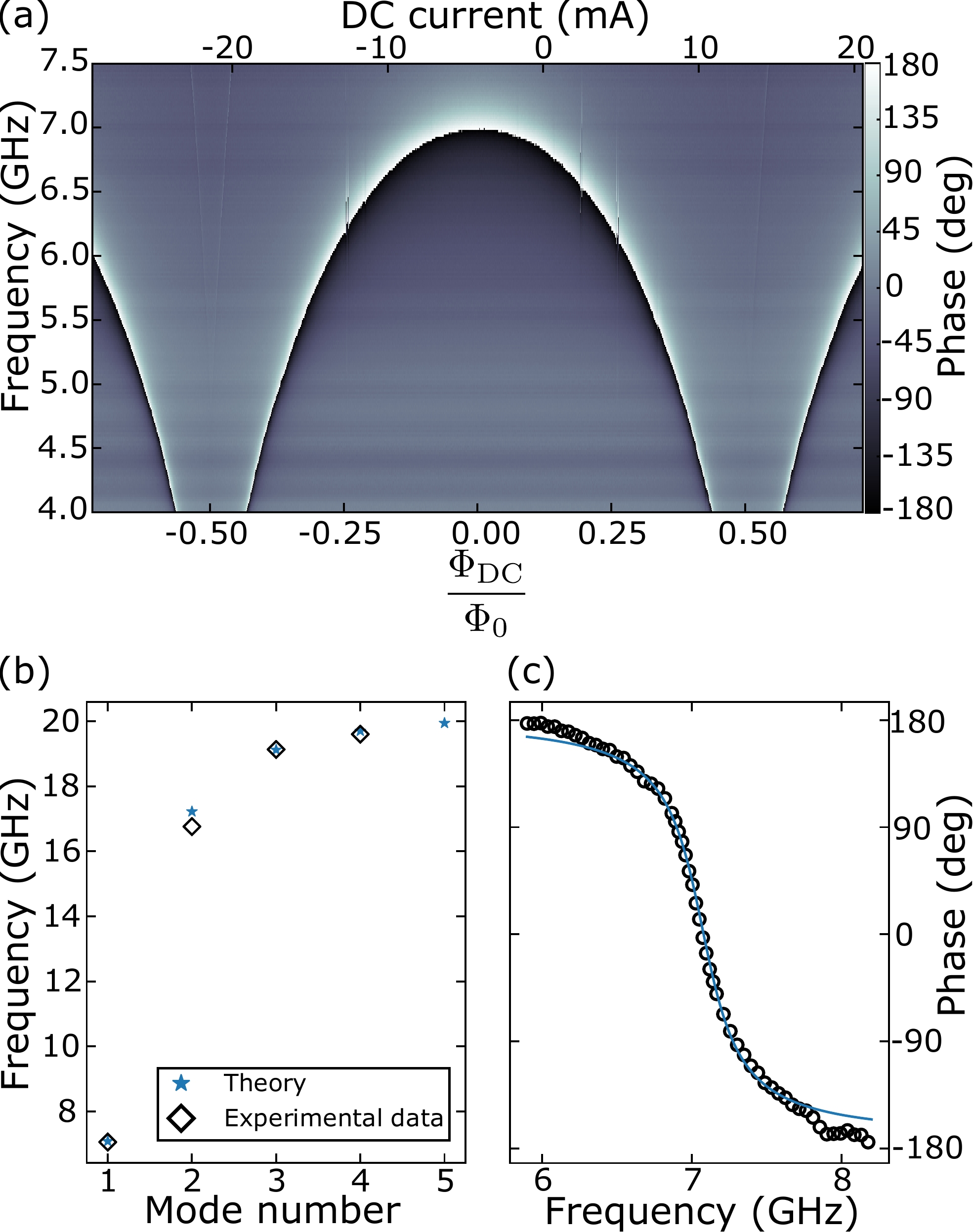}
\caption{
\textbf{a} DC flux modulation of the phase of the reflected signal. \textbf{b} Dispersion relation of the SQUID array for the first modes.  Blue stars are the solution of the matrix computation and black diamonds show experimental data. \textbf{c} Cut of \textbf{a} at zero magnetic flux, where the experimental data are fitted by an arctangent.}
\label{fig3}
\end{figure}

The device is measured using a conventional cryogenic microwave measurement setup (See Appendix~\ref{sec:setup}). The linear properties of the JPA are inferred by measuring the reflected phase of the microwave signal at zero flux and low power (\cref{fig3}.\textbf{c}). The fit of the phase shift yields $\omega_\text{exp}/2\pi=\SI{7.07}{\giga\hertz}$ and $Q_\text{exp}=19$. The resonant frequency of the JPA can be adjusted over a broad frequency range when flux-biasing the SQUID array (\cref{fig3}.\textbf{a}). We note the smooth behaviour of the device during flux tuning, despite the presence of the SQUID array. We attribute this stability to the absence of superconducting ground plane.
We can go one step further in the characterization of the device and obtain the dispersion relation of the array using two-tone spectroscopy~\cite{Anonymous:2012jo,Weissl:2015do}. The higher-order resonant frequencies of the device are presented in \cref{fig3}.\textbf{b}. Fitting these data using the circuit model presented in Section~\ref{nonlinearity}, we can determine $L_\text{J}$ and $a_0$ independently. Plugging these values in the effective model introduced before, we obtain the values $L_\text{eff}=\SI{21}{\nano\henry}$, $C_\text{eff}=\SI{24}{\femto\farad}$ and $K_\text{eff}=\SI{80}{\kilo\hertz}$. These values translate to an effective resonant frequency $\omega_\text{eff}/2\pi=\SI{7.08}{\giga\hertz}$ and an effective external quality factor $Q_\text{eff}=19$ in very good agreement with the measured values. We note that this external quality factor is much smaller than internal quality factors $Q_\text{int}\sim 10^4$ we measured in devices fabricated using the same procedure~\cite{krupko2018kerr}. This justifies that internal losses can be safely neglected in our model. In the next section, we will use the value of $K_\text{eff}$ to explain the measured gain, bandwidth and $\SI{1}{\decibel}$ compression point of the JPA without any free parameters.

\section{Gain and input saturation power}

\begin{figure}[h]
\includegraphics[width=\linewidth]{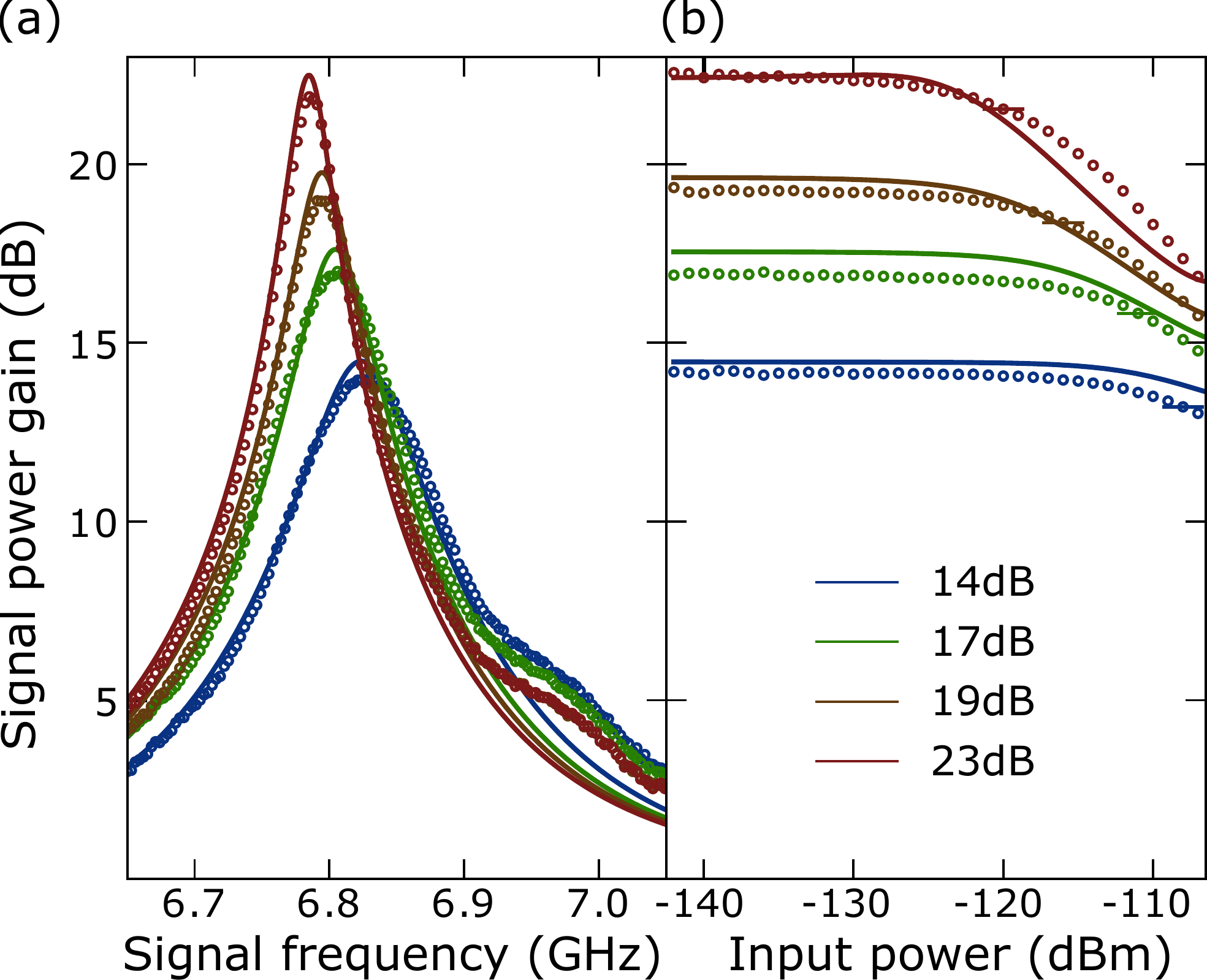}
\caption{
\textbf{a} Signal power gain versus signal frequency. Experimental (dots) and calculated (solid line) gain at four different pump powers and frequencies. The theoretical pump parameters are chosen as followed: the pump frequency is first set to the one used experimentally. The pump power is then set to maximise the gain at zero detuning ($\Delta$ = 0), as done experimentally. This leads to optimal pump biasing conditions, which can be visualised as the ridge on \cref{fig1}. These optimal conditions are (from low gain to high gain) : ($\SI{-81.65}{dBm}$, $\SI{6.83}{\giga\hertz}$), ($\SI{-81.12}{dBm}$, $\SI{6.80}{\giga\hertz}$), ($\SI{-80.83}{dBm}$, $\SI{6.79}{\giga\hertz}$), ($\SI{-80.57}{dBm}$, $\SI{6.78}{\giga\hertz}$). The bumps on the right tail of the experimental amplification curves are due to the normalization procedure and small losses at zero pump power. \textbf{b} Maximum gain as a function of the input power signal for the four same pump parameters. The pump powers for the theoretical plots have been shifted by up to $\pm\SI{0.03}{dBm}$ from the optimal pump power to account for the fact that a very small variation of pump power translates in a large variation of the gain as explained in the text. Such shifts are compatible with small drifts in the attenuation of the input line over the course of one day.}
\label{fig4}
\end{figure}

In \cref{fig4}.\textbf{a}, we present the gain of the amplifier versus frequency at various pump powers. We measure a $\SI{-3}{\decibel}$ bandwidth of $\Delta f=\SI{45}{\mega\hertz}$ at $\SI{20}{\decibel}$ of gain.  All these gain curves can be explained by~\cref{gain} using only the above-mentioned effective parameters. Interestingly it also provides an accurate calibration of the pump power at the JPA level and thus of the attenuation of the input line (see Appendix~\ref{sec:calib}). We note that our JPA can be flux tuned over a band of $\SI{900}{\mega\hertz}$ while reaching $G_\text{max}=\SI{20}{\decibel}$ as shown in Appendix~\ref{sec:flux}. Knowing the attenuation of the input line, the input saturation power of the JPA is quantified by measuring the maximum gain $G_\text{max}$ as a function of input power for different gains (\cref{fig4}.\textbf{b}). More specifically we measure a $\SI{1}{\decibel}$ compression point $P_\text{1dB}=-117\pm 1.4 \text{ dBm}$ at $\SI{20}{\decibel}$ of gain. This point corresponds to the input power at which the amplifier saturates and the gain is compressed by $\SI{1}{\decibel}$ from $G_\text{max}$.  Again we show a very good agreement between experiment and theory, without fitting parameters. To describe the saturation of the JPA, the number of signal photons inside the JPA must be taken into account while the pump is on. To do so, we add, in a self-consistent approach ~\cite{Eichler:2014iw}, the terms $2iK\langle a^{\dagger}a \rangle\alpha$ and $iK\langle a^{2} \rangle\alpha^{*}$ to the initial equation of motion of the intra-resonator field (see Appendix~\ref{sec:gain}). This correction to the total number of photons inside the cavity (pump, signal and idler), dependent on the signal power, allows the modelling of the amplifier saturation for a given set of pump frequencies and powers as plotted on \cref{fig4}.\textbf{b}. As will be explained in the next section, this saturation is very sensitive to the pump biasing conditions.

\section{Discussion}

\begin{figure}[htb]
\begin{center}
\includegraphics[width=\linewidth]{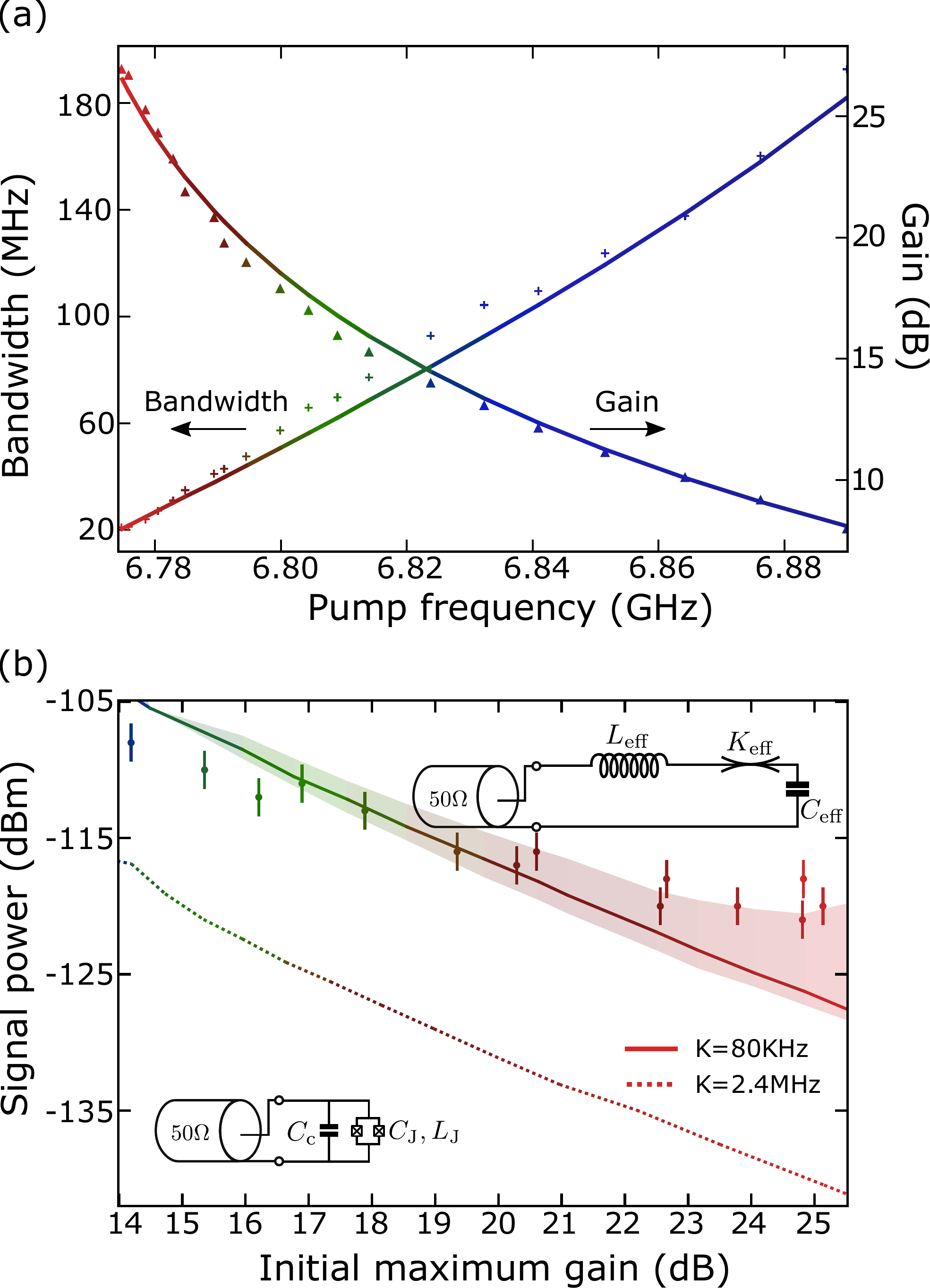}
\caption{Summary of the amplifier characteristics and agreement between experiment (dots) and theory for optimal pumping condition (line). \textbf{a} Maximum gain and $\SI{-3}{\decibel}$ bandwidth (obtained from a Lorentzian fit of the amplification curve) as a function of the pump frequency. The gain theoretical line follows the highlighted ridge shown on \cref{fig1}.\textbf{b}. \textbf{b} $\SI{1}{\decibel}$ compression point as a function of the initial maximum gain. The shaded area below (above) the theoretical curve shows the effect of a shift of $ + \SI{0.03}{dBm}$ ($ -\SI{0.03}{dBm}$) from the optimal pump power on the $\SI{1}{\decibel}$ compression point. The dashed line shows the $\SI{1}{\decibel}$ compression point of a single-SQUID JPA which would show the same bandwidth and operating frequency. } 
\label{fig5}
\end{center}
\end{figure}
To further illustrate the performance of our device and the predictive value of our model, we summarize three important figures of merit of our JPA in~\cref{fig5}. These are the maximum gain, the -3 dB bandwidth, and the 1 dB compression point. The maximum gain $G_\text{max}$ at low signal power and the corresponding $\SI{-3}{\decibel}$ bandwidth $\Delta f$ are measured for different pump powers (panels \textbf{a} and \textbf{b}). The gain-bandwidth product remains equal to $\SI{450}{\mega\hertz}$ over this pump power range, as expected from JPA theory. We now compare the 1 dB compression points measured at various gains to our theoretical predictions. Such a plot should be interpreted with great care since a very small deviation from optimal pump conditions can lead to variations of $P_{1dB}$ as reported previously~\cite{Liu:2017bi}. For example a pump power variation of $0.03\ \text{dBm}$ leads to a change of up to $\SI{3}{dBm}$ in $P_\text{1dB}$, as illustrated by the shaded area of \cref{fig5}.\textbf{c}. To illustrate the advantage of using SQUID arrays, we also plot what would be $P_\text{1dB}$ for a single SQUID JPA as reported in various papers~\cite{Hatridge:2011dh,Mutus:2013iw}. To ensure a meaningful comparison we chose the parameters of this single SQUID JPA so that it displays the same working frequency ($\omega_\text{exp}/2\pi=\SI{7.07}{\giga\hertz}$) and bandwidth ($Q_\text{exp}=19$) as our array JPA. The self-Kerr coefficient of such JPA would be $K_\text{single}/2\pi=\SI{2.4}{\mega\hertz}$ (to be compared to $K_\text{eff}/2\pi=\SI{80}{\kilo\hertz}$). This translates into $P_\text{1dB,single}= -131\ \text{dBm}$ at $G_\text{max}=\SI{20}{\decibel}$ compared to $P_\text{1dB,array}=-116\text{ dBm}$ for our array JPA. This $15 \text{ dBm}$ difference reflects directly the ratio of self-Kerr coefficients since $P_\text{1dB}$ scales as $\kappa_\text{eff}/\abs{K_\text{eff}}$ as explained previously. This illustrates the key advantage of using arrays to fabricate high-performance JPAs. Finally we would like to discuss the data/theory agreement. According to our microscopic model $P_\text{1dB}$ should be $-116\text{ dBm}$, while we measured $-117\pm1.4\text{ dBm}$. This good agreement confirms that adding two terms to the equation of motion of the intra-resonator field is enough to explain the saturation effect observed in our JPA. From a physical point view, the effect of these terms is two-fold. First the bare frequency of the JPA $\omega_\text{eff}$ becomes dependent on the number of signal photons, similarly to the ac-Stark shift effect. Second the number of pump photons inside the JPA depends as well on the number of signal photons; an effect known as pump depletion in parametric amplifiers theory.

\section{Conclusion}

We designed and measured a Josephson parametric amplifier made of 80 SQUIDs. This device relies on a single-step, all-aluminium fabrication process, easily reproducible in a research-grade clean-room. We showed that the number of SQUIDs in the array has a direct and predictable impact on the nonlinearity, which is directly linked to the saturation power of the amplifier. The circuit model we used gives a very good agreement with the experimental data, without fitting parameters. Improvements could be obtained by bringing the Josephson inductance down to $L_\text{J} \approx L_\text{stray}$. Setting $L_\text{J}$ to \SI{40}{\nano\henry}, just above $L_\text{stray}$, adjusting $C_\text{out}$ to $\SI{50}{\femto\farad}$ and the total number of SQUIDs to $N=150$, would lead to a JPA with a bare resonant frequency $f_0=\SI{7.45}{\giga\hertz}$ and external quality factor $Q_\text{e}=9$. According to our model, this JPA would display for \SI{20}{\decibel} signal gain, a bandwidth of \SI{95}{\mega\hertz} and a \SI{1}{\decibel} compression point of $-102\text{dBm}$. A pump power of $-66\ \text{dBm}$ would be necessary to operate a JPA with these figures of merit. This value is comparable to what was reported for Josephson Traveling Wave Parametric Amplifiers (JTWPA)~\cite{macklin2015near,white2015traveling} and, as such, should not be a concern. We would like to stress that these estimates cannot be strictly quantitative since the approximation described in \cref{sample} ($L_\text{stray}\ll L_\text{J}$) does not hold anymore. Theory should be further developed to account for the effect of these stray inductances. Further developments that could be applied to this SQUID array JPA include input impedance engineering to improve the performance of the device~\cite{Mutus:2014dd,Roy:2015ky} or band engineering to bring in new capabilities such as non-degenerate~\cite{Bergeal:2010iu} or multi-mode parametric amplification~\cite{Simoen:2015by}.

\section*{Acknowledgements}

The authors would like to thank W. Wernsdorfer, E. Eyraud,
F. Balestro and T. Meunier for early support with the experimental setup. Very
fruitful discussions with F. W. Hekking and D. Basko and I. Takmakov are strongly
acknowledged. This research was supported in part by the International Centre for Theoretical Sciences (ICTS) 
during a visit for participating in the program - Open Quantum Systems (Code: ICTS/Prog-oqs2017/2017/07). This research was supported by the ANR under contracts CLOUD
(project number ANR-16-CE24-0005). J.P.M. acknowledges
support from the Laboratoire d\textquoteright excellence LANEF in Grenoble
(ANR-10-LABX-51-01). R.D. and S.L. acknowledge support from the CFM
foundation.\\

\appendix

\section{Experimental setup\label{sec:setup}}

\begin{figure}[h]
\includegraphics[width=\linewidth]{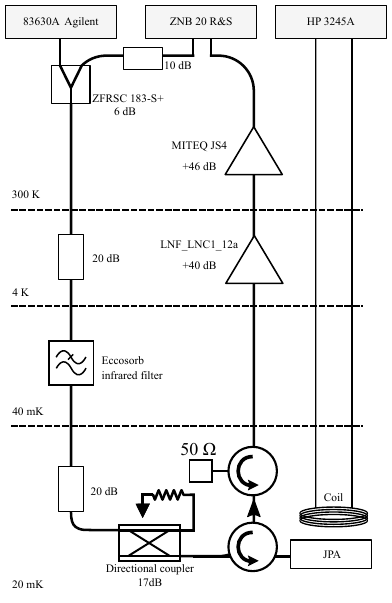}
\caption{Experimental setup}
\label{experimental_setup}
\end{figure}

The full measurement setup is shown in \cref{experimental_setup}. The device was placed in a dilution refrigerator at a base temperature of \SI{20}{\milli\kelvin}, and the transmission measurements were performed using a Vector Network Analyzer (VNA).  An additional microwave source was used for two-tone measurements while a global magnetic field was applied via an external superconducting coil. The output line included one isolator  at \SI{20}{\milli\kelvin}, a HEMT amplifier at \SI{4}{\kelvin} and room temperature amplifiers. The input line was attenuated at various stages, including a home-made filter that prevents stray-radiations from reaching the sample. We adopted a coaxial geometry with a dissipative dielectric (reference RS-4050 from resin systems company). The bandwidth of the measurement setup goes from \SIrange{4}{13}{\giga\hertz}. 

\section{Diluting the non-linearity in a Josephson array \label{sec:nonlin}}

In this appendix, we briefly demonstrate the effect of an array of Josephson junctions on the effective nonlinearity. For the sake of simplicity, we don't take into account the propagation effects. First, we start to derive an expression for the nonlinearity for a single junction whose Hamiltonian is

\begin{equation}
\hat{H} = \frac{\hat{Q}^2}{2C} + E_\text{J}\cos{\hat{\varphi}}
\end{equation}

where $\hat{Q}$ is the charge operator and $\hat{\varphi}$ is its conjugate such as $[\hat{Q},\hat{\varphi}]=-2ie$, with $e$ the positive charge of the electron. $C$ is the capacitance of the junction, $E_\text{C}=q/2C$ and $E_\text{J}$ are the charging and the Josephson energy of the junction, respectively, defining the resonance frequency of the junction $\hbar\omega_\text{o}=\sqrt{8E_\text{J}E_\text{C}}$. Under the assumption that the phase fluctuations are small, the cosine potential is developed up to the fourth order $E_\text{J}\cos{\hat{\varphi}} = E_\text{J} - E_\text{J}\hat{\varphi}^2/2! + E_\text{J}\hat{\varphi}^4/4! + o(\hat{\varphi}^4) $. Then $\hat{q}$ and $\hat{\varphi}$ are written as a function of creation and annihilation operators:

\begin{subequations}
\begin{align}
\hat{q} &= q_\text{zpf}(\hat{a}^{\dagger} + a) \\
\hat{\varphi} &= i\varphi_\text{zpf}(\hat{a}^{\dagger} - a)  
\end{align}
\end{subequations}

Where the zpf stands for zero point fluctuations, with $q_\text{zpf}=\sqrt{\hbar\omega C/2}$ and $\varphi_\text{zpf}=\sqrt{\hbar/2C\omega}\propto(E_\text{C}/E_\text{J})^\frac{1}{4}$.
The Hamiltonian is rewritten as: 

\begin{equation}
\hat{H} = \frac{1}{2C}q_\text{zpf}^2(\hat{a}^{\dagger} + a)^2 + \frac{E_\text{J}}{2}\varphi_\text{zpf}^2(\hat{a}^{\dagger} - a)^2 - \frac{E_\text{J}}{4!}\varphi_\text{zpf}^4(\hat{a}^{\dagger} - a)^4
\label{hamil_aa}
\end{equation}

The last term on the right hand side of \cref{hamil_aa} is the nonlinearity linked to the self-Kerr coefficient by $K=E_\text{J}\varphi_\text{zpf}^4/4!\hbar \propto E_\text{C}/\hbar$~\cite{Koch:2007gz}.
 
Let's consider now the $N$ series junctions case. The Josephson potential is written $NE_\text{J}\cos{\frac{\hat{\varphi}}{N}}$, under the assumption that the phase-drop across the chain is equally divided across each junction. The potential is once again developed to the fourth order $NE_\text{J}\cos{\frac{\hat{\varphi}}{N}} =   NE_\text{J} - NE_\text{J}\hat{\varphi}^2/2N^2 + NE_\text{J}\hat{\varphi}^4/4!N^4$. The second term is simplified in $E_\text{J}\hat{\varphi}^2/2N$, and a new Josephson energy is defined as $E_\text{J}^* = E_\text{J}/N$. Considering that we want to keep the same resonance frequency, it leads to

\begin{equation}
\hbar\omega_0 = \sqrt{8E_\text{C}E_\text{J}} = \sqrt{8E_\text{C}^*E_\text{J}^*}
\end{equation}

This condition gives $E_\text{C}^* = NE_\text{C}$ and leads to $\varphi_{ZPF}^* \propto (E_\text{C}^*/E_\text{J}^*)^{\frac{1}{4}} \propto N^{\frac{1}{2}}(E_\text{C}/E_\text{J})^{\frac{1}{4}} \sim N^{\frac{1}{2}}\varphi_{ZPF}$. Now the new Kerr term $K^*$ is :

\begin{equation}
\hbar K^* = \frac{NE_\text{J}}{4!}\frac{{\varphi_{ZPF}^*}^4}{N^4} = \frac{E_\text{J}}{4!}\varphi_{ZPF}^4\frac{1}{N} = \frac{\hbar K}{N}
\end{equation}

We demonstrated that with an array of $N$ Josephson junctions, while keeping the same resonance frequency and under the assumption that each junction is equally phase-biased in the array, the non-linearity is divided by $N$ compared to the single junction case. This calculation is only meant to give a qualitative estimation since in a real device propagating effects have to be accounted for, as explained in the main text.

\section{Flux Tunability\label{sec:flux}}

\begin{figure}[h]
\includegraphics[width=\linewidth]{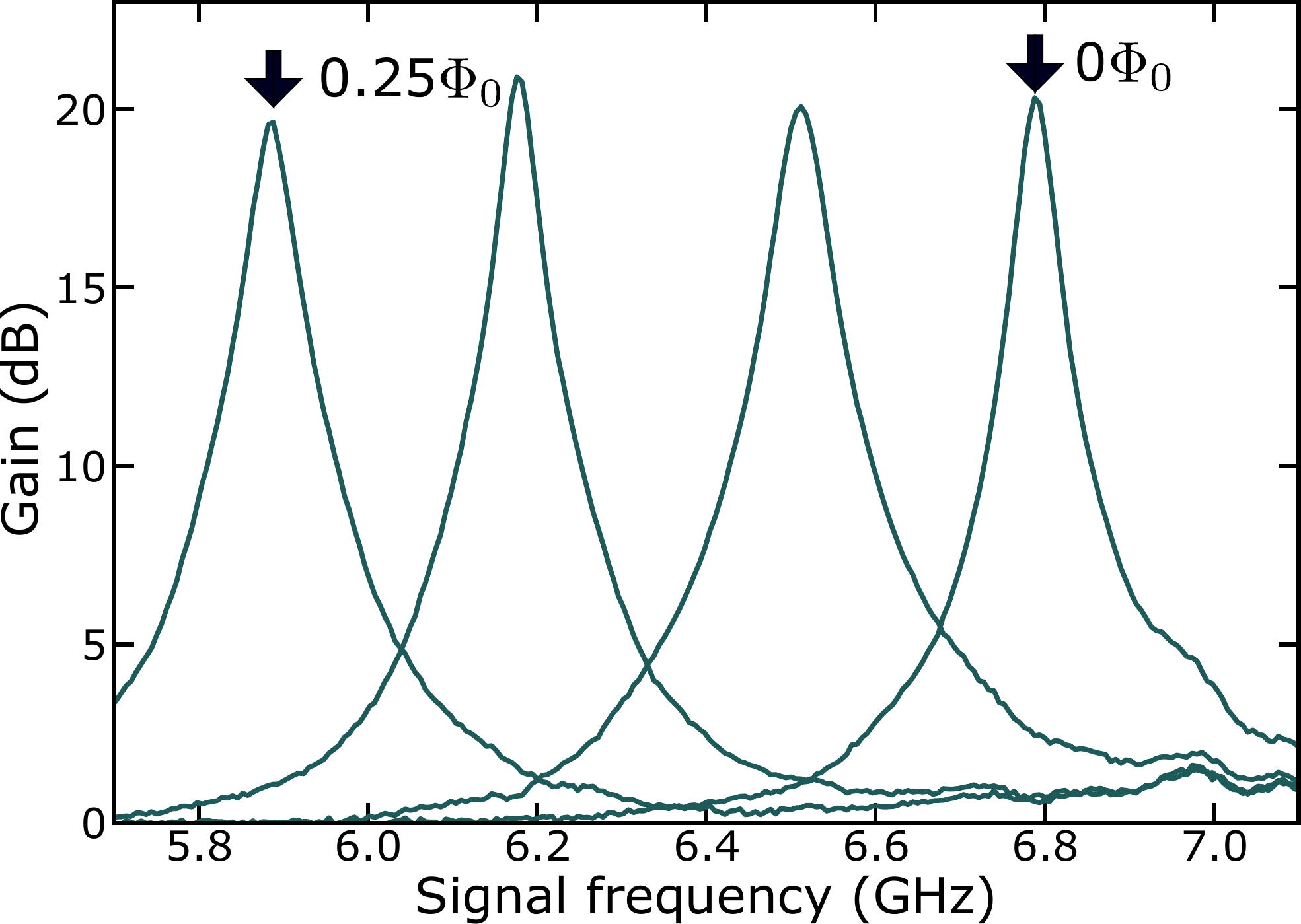}
\caption{Gain of the amplifier for different flux bias configurations.}
\label{flux}
\end{figure}

In this section, we show the frequency range on which the amplification is possible while flux tuning the array. In the main text, we presented the flux tunability of the JPA by showing the 2$\pi$ phase shift from $\SI{7}{\giga\hertz}$ to $\SI{4}{\giga\hertz}$, which corresponds to the bare frequency of the array and the lower bound of the circulator, respectively. Nonetheless, this frequency window does not correspond to the band on which amplification is reachable. We arbitrarily define the range of effective tunability as the range where $\SI{20}{\decibel}$ signal gain can be observed. We show in \cref{flux} signal gain for different DC flux biasing. We could obtain clear amplification from $\sim \SI{6.8}{\giga\hertz}$ ($\Phi=0\Phi_0$) to $\sim \SI{5.9}{\giga\hertz}$ ($\Phi\approx0.25\Phi_0$). At lower frequencies, the critical current of the SQUID decreased too much compared to the pump current necessary to achieve $\SI{20}{\decibel}$ gain.

\section{Remote Ground Model \label{sec:Remote}}
In~\cref{tab}, we compare 4 SQUID array models to experimental results and justify our choice to use the remote ground model, as described in section~\ref{nonlinearity} of the main text. The simplest way to model such an array is to only consider the Josephson inductances $L_\text{J}$ and the capacitances $C_\text{out}$ and $C_\text{s}$, without taking into account the capacitive effect between the SQUIDs and the ground. With this over-simplified model, we can choose $L_\text{J}$ to obtain the right resonant frequency but the value of the quality factor is then wrong and vice-versa. Moreover no propagating effects can be described and thus the dispersion relation cannot be reproduced. The standard model using a ground capacitance $C_\text{g}$ for each elementary cell (local screening of the charge $Q_\text{n}$ by the ground) can reproduce the dispersion relation, but the values of the external quality factor and resonant frequency are less accurate. Finally, the remote ground model reproduces well the dispersion relation and returns an effective quality factor $Q_\text{eff}$ closer to the experimental one. An intuitive way to estimate the effective parameters of the circuit without going through the matrix computation shown in the main text is to take $C_\text{out}$, $C_\text{J}$ and $C_\text{s}$ equal to zero and consider the array as a simple transmission line. Then each unitary cell has an inductance $L_\text{J}$, a ground capacitance $C_\text{g}$ and an impedance $Z_{\text{TL}}=\sqrt{L_\text{J}/C_\text{g}}$. In that case, there is a direct mapping between a $\frac{\lambda}{4}$ resonator and an effective $LC$ series resonator close to resonance~\cite{pozar2009microwave}. We can define an effective inductance and an effective capacitance as:

\begin{subequations}
\begin{align}
L_\text{eff,TL} &= \frac{\pi Z_\text{TL}}{4\omega_0} \\
C_\text{eff,TL} &= \frac{1}{\omega_0^{2} L_\text{eff,TL}}  
\end{align}
\end{subequations}

By setting the first resonance to $\omega_0/2\pi$=\SI{7.07}{\giga\hertz}, $L_\text{J}$ and $C_\text{g}$ to the values inferred from the measured dispersion relation (\SI{165}{\pico\henry} and \SI{0.3}{\femto\farad} respectively), we find a characteristic impedance $Z_\text{c}=\sqrt{L_\text{eff,TL}/C_\text{eff,TL}}=\SI{600}{\ohm}$. The external quality factor is then given by $Q_\text{eff,TL}=Z_\text{c}/Z_\text{0}=12$. This over simplified model fails to reproduce, at the same time, the experimental values of the resonant frequency and the external quality factor. In the following table, we present the different effective parameters found ($\omega_\text{eff}$ , $Q_\text{eff}$ and $K_\text{eff}$ ) using the various models as well as the experimental values.

\begin{table}[ht]
\begin{tabular}{|C{2cm}|C{2cm}|C{1cm}|C{2cm}|}
\hline  \rowcolor{lightgray}   & $f_\text{eff}$ (GHz)  & $Q_\text{eff}$ & $K_\text{eff}$ (KHz)\\
\hline  Experimental & 7.07 & 19 & x \\
\hline  No ground Capacitance & 7.95  & 13 & 98  \\
\hline  Ground capacitance & 7.05  & 21 & 80  \\
\hline  Remote ground & 7.08 & 19 & 80  \\
\hline  $\frac{\lambda}{4}$ resonator & 7.07 & 12 & x  \\
\hline
\end{tabular}
\caption{Comparison of the effective parameters between three models of the SQUID array and the parameters found experimentally.}
\label{tab}
\end{table}

\section{Derivation of the gain \label{sec:gain}}

In this section, we detail the derivation to obtain expression of the gain, as a function of the pump and signal parameters. As in the main text, we start describing the circuit with the Hamiltonian of a non-linear resonator with annihilation and creation operators:

\begin{equation}
H_\text{JPA}=\hbar\omega_\text{eff}A^\dagger A+\hbar\frac{K_\text{eff}}{2}\left (A^\dagger\right)^2 A^2
\end{equation}   

The dynamics of the circuit is described with the conventional input-output theory

\begin{equation}
\dot{A} = -i\omega_\text{eff}A-iK_\text{eff}A^\dagger AA - \frac{\kappa_\text{eff}}{2}A+\sqrt{\kappa_\text{eff}} A_\text{in}
\label{EOM}
\end{equation}

We neglect the internal losses as they are much smaller than the coupling rate $\kappa_\text{eff}$. In this derivation, the intra-resonator field $A(t)=(\alpha + a(t))e^{i\omega t}$ is decomposed in two components: a strong, classical field $\alpha e^{i\omega t}$ called pump and a weak, quantum field $a(t)$ which we refer to as the signal. First, \cref{EOM} is considered only with the pump field $A(t) = \alpha(t)e^{i\omega t}$. We multiply both sides with their complex conjugate, leading to

\begin{equation}
1 = (\delta^2 + \frac{1}{4})n - 2\delta\xi_\text{$\alpha$} n^2 + \xi_\text{$\alpha$}^2n^3
\label{EOM_simpl}
\end{equation}

where $\delta=(\omega_\text{p} - \omega_\text{eff})/\kappa_\text{eff}$ is the detuning between the pump and the bare frequency of the resonator, $\tilde{\alpha}_\text{in}=\alpha_\text{in}/\sqrt{\kappa_\text{eff}}$ is the dimensionless drive amplitude, $\xi_\text{$\alpha$}=|\tilde{\alpha}_\text{in}|^2K_\text{eff}/\kappa_\text{eff}$, is the pump strength and finally $n=|\alpha|^2/|\tilde{\alpha}_\text{in}|^2$ is the mean number of pump photons inside the non-linear resonator. We numerically solve this equation which is cubic in $n$ to determine the number of pump photon as a function of the pump power and pump frequency. Once this equation solved, the signal tone is added ($A(t) = \alpha$(t) + a(t)), \cref{EOM} is linearized for the weak quantum signal, only the first order terms in $a(t)$ are kept:

\begin{equation}
\begin{split}
\dot{a}(t) = i(\omega_\text{p} - \omega_\text{eff} - 2K_\text{eff}|\alpha|^{2} + i\frac{\kappa_\text{eff}}{2})a(t)  \\
- iK_\text{eff}\alpha^2a^{\dagger}(t) + \sqrt{\kappa_\text{eff}} a_\text{in}
\label{EOM_lin}
\end{split}
\end{equation}

To solve \cref{EOM_lin}, $a(t)$ is decomposed into its Fourier components since \cref{EOM_lin} is linear in $a(t)$:

\begin{equation}
a(t) = \frac{\kappa_\text{eff}}{\sqrt{2\pi}}\int^{\infty}_{-\infty}{a_{\Delta}d\Delta e^{-i\Delta\kappa_\text{eff} t}}
\label{Fourier}
\end{equation}

where $\Delta = (\omega_s - \omega_p)/\kappa_\text{eff}$ is the dimensionless detuning between the pump and the signal. \cref{EOM_lin} can be rewritten as a function of the Fourier components of $a(t)$, using the parameters defined before:

\begin{equation}
0 = (i(\delta - 2\xi_\text{$\alpha$}n + \Delta) - \frac{1}{2})a_{\Delta} -i\xi_\text{$\alpha$} n e^{2i\phi}a^{\dagger}_{-\Delta} + \tilde{a}_{in,\Delta}
\label{EOM_lin_simple}
\end{equation}

Since \cref{EOM_lin} mixes $a_{\Delta}$ and its conjugate $a^{\dagger}_{-\Delta}$, the conjugate of \cref{EOM_lin} has to be accounted for to have a full expression of $a_{\Delta}$ as a function of the dimensionless input $\tilde{a}_{in,\Delta} = {a}_{in,\Delta} / \sqrt{\kappa_\text{eff}} $ and its conjugate $\tilde{a}_{in,\Delta}^{\dagger} = {a}_{in,\Delta}^{\dagger} / \sqrt{\kappa_\text{eff}}$. This leads to a set of two equations linking input and output, which can be written as a matrix equation:

\begin{widetext}
\begin{equation}
\colvec{2}{\tilde{a}_{in,\Delta}}{\tilde{a}_{in,-\Delta}^{\dagger}} = 
		\begin{pmatrix}
		   i(-\delta + 2\xi_\text{$\alpha$} n - \Delta) + \frac{1}{2}  &  i\xi_\text{$\alpha$} n e^{-i2\phi} \\  
		   -i\xi_\text{$\alpha$} n e^{-i2\phi} &   i(\delta - 2\xi_\text{$\alpha$} n - \Delta) + \frac{1}{2}  
	   \end{pmatrix}\colvec{2}{{a}_{\Delta}}{{a}_{-\Delta}^{\dagger}}
\label{EOM_matrix}
\end{equation}
\end{widetext}

By inverting the matrix, we can have access to the expression of ${a}_{\Delta}$ and ${a}_{-\Delta}^{\dagger}$ as a function of the input field $\tilde{a}_{in,\Delta}$ and $\tilde{a}_{in,-\Delta}^{\dagger}$ and the pump parameter $\delta$, $\xi_\text{$\alpha$}$, $n$ and $\Delta$:

\begin{widetext}
\begin{equation}
\begin{split}
a_{\Delta} = \frac{i(\delta - 2\xi_\text{$\alpha$} n - \Delta) + \frac{1}{2}}{(i\Delta - (\frac{1}{2} - \sqrt{ (\xi_\text{$\alpha$} n)^2 - (\delta - 2\xi_\text{$\alpha$} n)^2} ))(i\Delta - (\frac{1}{2} + \sqrt{ (\xi_\text{$\alpha$} n)^2 - (\delta - 2\xi_\text{$\alpha$} n)^2} ))}\tilde{a}_{in,\Delta} \\
+ \frac{-i\xi_\text{$\alpha$} n e^{2i\phi}}{(i\Delta - (\frac{1}{2} - \sqrt{ (\xi_\text{$\alpha$} n)^2 - (\delta - 2\xi_\text{$\alpha$} n)^2} ))(i\Delta - (\frac{1}{2} + \sqrt{ (\xi_\text{$\alpha$} n)^2 - (\delta - 2\xi_\text{$\alpha$} n)^2} ))}\tilde{a}_{in,-\Delta}^{\dagger}
\label{a}
\end{split}
\end{equation}
\end{widetext}

Finally, we can link the intra-cavity field $a_{\Delta} $ with the output field with the boundary condition :

\begin{equation}
a_{\text{out,}\Delta} = \sqrt{a_\Delta} - a_{\text{in},\Delta}
\label{boundary}
\end{equation}
leading to an expression linking the output field to the input field.
\begin{equation}
\begin{split}
a_{out,\Delta} = -1 + \frac{i(\delta - 2\xi_\text{$\alpha$} n - \Delta) + \frac{1}{2}}{(i\Delta - \lambda_{-})(i\Delta - \lambda_{+})}\tilde{a}_{in,\Delta} \\
+ \frac{-i\xi_\text{$\alpha$} n e^{2i\phi}}{(i\Delta - \lambda_{-})(i\Delta - \lambda_{+})}\tilde{a}_{in,-\Delta}^{\dagger}  
\label{a simple}
\end{split}
\end{equation} 
with $\lambda_{\pm}=\frac{1}{2} \pm \sqrt{((\xi_\text{$\alpha$} n)^2 - (\delta - 2\xi_\text{$\alpha$} n))}$. Signal and idler gains are defined as the ratio between output field and input field at $\Delta$ and $-\Delta$ respectively. We can define the signal gain as the prefactor of $\tilde{a}_{in,\Delta}$ in \cref{a simple}, as in the main text. The maximum gain at zero detuning between the pump and the signal ($\Delta=0$) as a function of the pump power and pump frequency is plotted in \cref{fig1} with our JPA parameters.

\begin{figure}
\includegraphics[width=\linewidth]{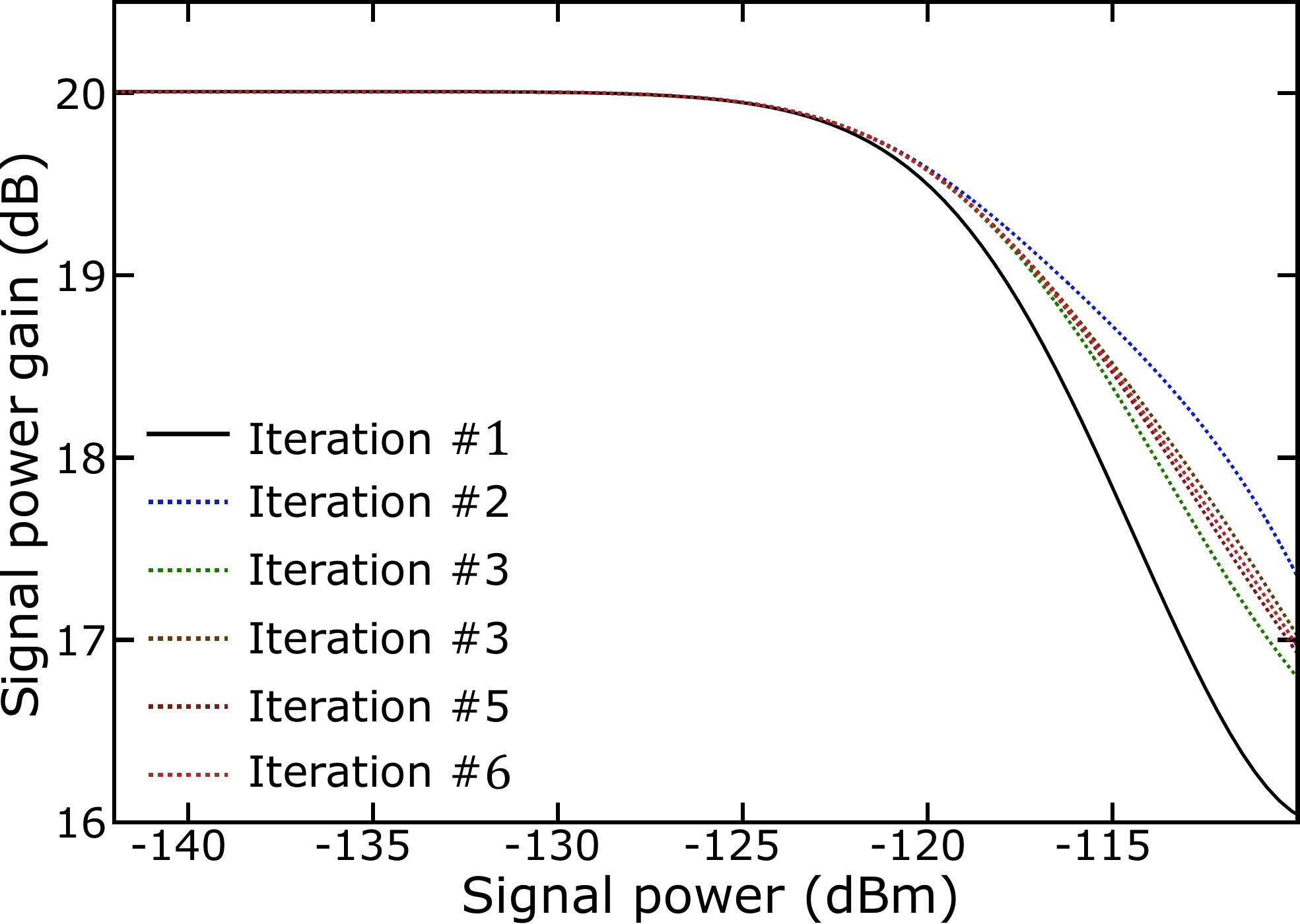}
\caption{Maximum gain as a function of the input signal power where the actual number of signal photons inside the resonator is being iteratively computed five times.}
\label{conv}
\end{figure}

Gain versus signal-pump detuning is plotted in \cref{fig1}. Theory and experimental points are obtained following the same protocol: for a given pump frequency (the same for theory and experiment), the pump power is set to obtain the maximal signal gain at $\Delta=0$, with no regard to other parameters. By comparing the experimental, injected pump power and the theoretical, expected one, we were able to make a calibration of the input line (see Appendix  \ref{sec:calib}). Up to now the mean number of signal photons was set to 0. To go further in our understanding of the JPA, the saturation was modelled by taking into account the number of signal photons inside the non-linear cavity. To take into account this number, the terms $2iK\langle a^{\dagger}a \rangle\alpha$ and $iK\langle a^{2} \rangle\alpha^{*}$, previously neglected, are now added, to \cref{EOM}. Following the same mathematical steps with these two new terms we obtain a new cubic equation in $n$, the mean number of photon inside the cavity, now depending not only on pump parameters but also on the input signal power:

\begin{widetext}
\begin{equation}
\begin{split}
1=n[ \delta^2 + \frac{1}{4} - 4\xi_a\delta + 5\xi_\text{a}^2 + \xi_a((2\xi_\text{a}^2 + \delta)\cos{2\Delta\phi} + \frac{1}{2}\sin{2\Delta\phi}  ) ] \\
+ n^2[ -2\delta\xi_\text{$\alpha$} + 4\xi_\text{$\alpha$}\xi_a + \xi_\text{$\alpha$}\xi_a\cos{2\Delta\phi}] + n^3\xi_\text{$\alpha$}^2
\label{new_n}
\end{split}
\end{equation}     
\end{widetext}

where $\xi_\text{a}=\frac{K_\text{eff}n_\text{a}}{\kappa_\text{eff}}$ is the normalized signal strength, $n_\text{a}$ is the mean photon number created by the input signal power $n_\text{a} = \langle a^{\dagger}_{\Delta}a_{\Delta} \rangle + \langle a^{\dagger}_{-\Delta}a_{-\Delta} \rangle$ and $\Delta\phi$ the phase difference between the pump and the signal. To consider the terms depending on $\Delta\phi$ (operating in phase sensitive mode) the difference between the signal frequency and the pump frequency has to be smaller than IF bandwidth. Otherwise, these terms average to 0. Given the IF bandwidth $\kappa_\text{if}$=$\SI{10}{\hertz}$ used in our experiment, these terms can be safely neglected. We link $n_\text{a}$ to the input power using \cref{a}, giving the intra-cavity field as a function of the input power. Moreover we neglect the idler input power $\tilde{a}_{in,-\Delta}^{\dagger}$ and the associated vacuum fluctuations, which leads to:

\begin{subequations}
\begin{align}
a_{\Delta} = \frac{i(\delta - 2\xi_\text{$\alpha$} n - \Delta) + \frac{1}{2}}{(i\Delta - \lambda_{-} )(i\Delta - \lambda_{+} )}\tilde{a}_{in,\Delta} \\
a_{-\Delta}^{\dagger} = \frac{i\xi_\text{$\alpha$}n e^{2i\phi}}{(i\Delta - \lambda_{-} )(i\Delta - \lambda_{+} )}\tilde{a}_{in,\Delta} 
\end{align}
\end{subequations}
\label{coef_conv}
 
where $|\tilde{a}_{in,\Delta}|^2$=$P_\text{signal}/\hbar\omega_s\kappa_\text{eff}$. Using this formula, we compute the photon number inside the cavity while taking into account the signal power and the Kerr shift induced by the signal power. We set the pump parameters to reach the optimal maximum gain (as explained in the previous paragraph), while $\Delta=0$. We observe a decrease in the gain as the signal power increases, as shown in the main text in \cref{fig3}.\textbf{b}. Since this mean-field approach needs to be solved self-consistently, once this new pump photon number $n$ (solution of \cref{new_n} ) is obtained, $n$ has to be plugged again in \cref{a} to get the actual number of signal photon inside the cavity, which will affect the pump photon number inside the cavity and so on. We iterate this loop several times and show on \cref{conv} that the saturation process converges after four iterations. In this study, input power of the quantum fluctuations have not been taken into consideration in the calculation of the mean photon number $n_\text{a}$. Care has been taken to ensure that saturation of the amplifier starts for $n_\text{a}$ greater than one, which mean that input power of the quantum fluctuations is negligible in the saturation. Nonetheless, in the calculation of the \SI{1}{\decibel} compression point for a Kerr constant equals to \SI{2.4}{\mega\hertz} in the main text, saturation occurs for $n_\text{a}$ smaller than one for gain greater than \SI{20}{\decibel}. This means that quantum fluctuations themselves saturate the amplifier. A model taking into account these fluctuations in a self-consistent way is beyond the scope of this paper. 

\section{Calibration of the input line\label{sec:calib}}

\begin{figure}
\includegraphics[width=\linewidth]{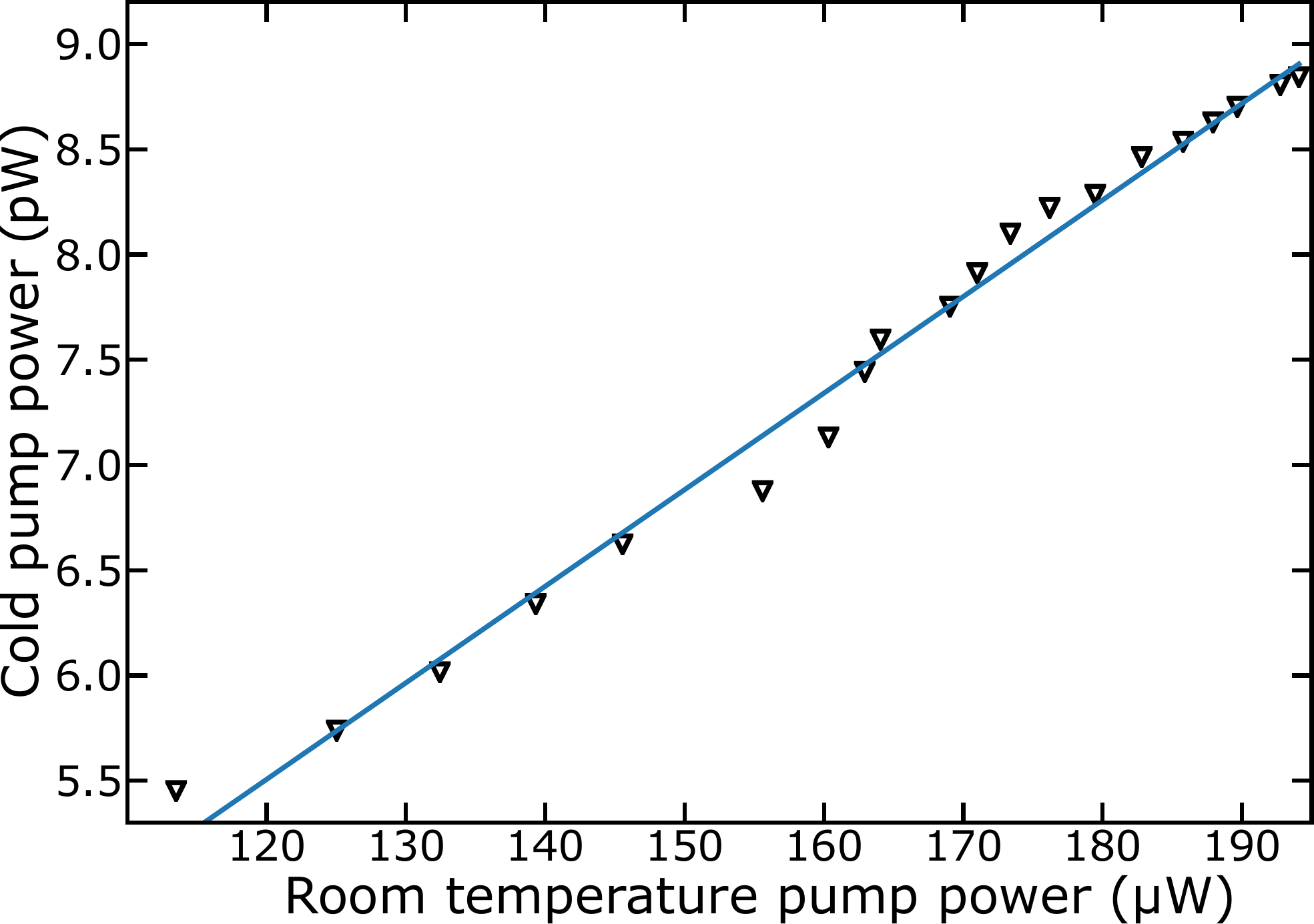}
\caption{Power at the input of the JPA versus room temperature power.}
\label{PvsP}
\end{figure}
In this section, we detail how the calibration of the input was done. To do so, we plugged our JPA characteristics (resonant frequency, bare bandwidth and nonlinearity) in the model detailed in the last section. As explained, we compared the amplification process (by sweeping the signal frequency around the pump frequency) between theory and experimental data for several pump sets. To choose a set of pump parameters, we followed the same protocol for both theory and experimental data: once the pump frequency was fixed, we looked for the pump power leading to the greater gain at $\Delta=0$. Considering the predictability of the model (see Fig. 3.\textbf{a} and Fig. 3.\textbf{b}), we compared experimental, room temperature power at the output of the pump source with theoretical power at the input of the JPA. We plotted on \cref{PvsP} points whose ordinate is the theoretical power and abscissa is the experimental one. We fitted these points with a linear relation, whose the intercept is 0 and the slope is $\SI{-73.4}{\decibel}$. We set the total attenuation between the pump output and the JPA input to $\SI{-73.0}{\decibel}$. This calibration is consistent with the $\SI{63}{\decibel}$ of discrete attenuators between the pump output and the JPA, and with $\SI{7}{\decibel\per\m}\times\SI{1.5}{m}+\SI{3}{\decibel\per\m}\times\SI{0.4}{m}=\SI{11.7}{\decibel}$ of attenuation across the different cables (see \cref{experimental_setup}). Therefore there is a discrepancy of \SI{1.4}{\decibel} between our calibration based on the self Kerr effect and our estimation based on the characteristics of the cables. We use this difference to set the size of the error bars in our measurements.

\section{Noise properties of the amplifier \label{sec:noise}}

\begin{figure}	
\begin{center}
\includegraphics[width=\linewidth]{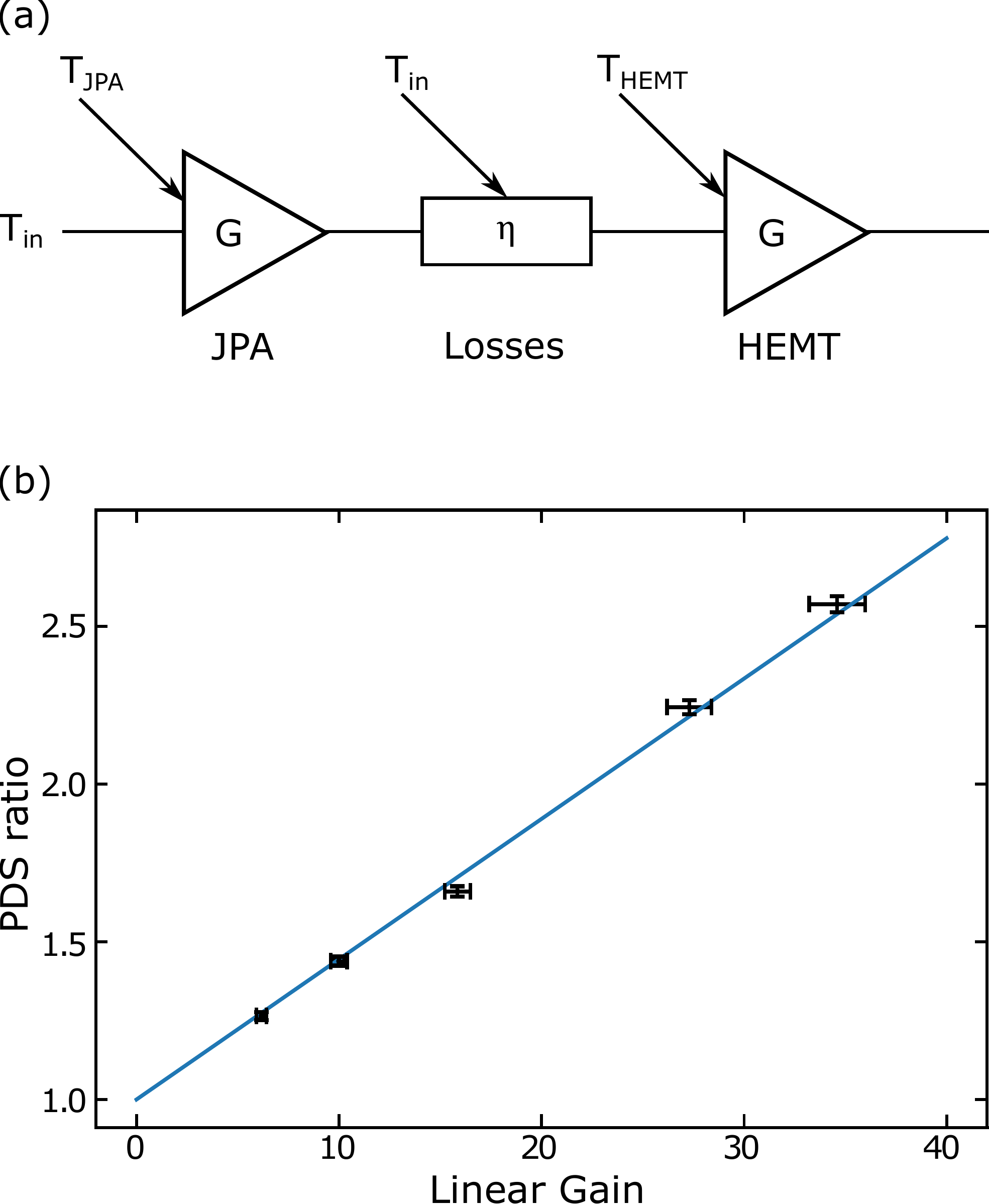}
\caption{(a) Simplified diagram of the amplification chain (b) $PSD$ ratio as a function of the linear JPA gain.}
\label{fig:JPAnoise}
\end{center}
\end{figure}
We discuss here the noise properties of a JPA nominally identical to the one presented in this paper. The analysis is closed to the one described in appendix B of \textit{Lin et al.} paper~\cite{Lin:2013cj}. We measure the power spectral density at the output of the whole measurement chain when the JPA is off and compare it to the case where it is operated with finite gain. This measurement chain consists in a JPA followed by a HEMT amplifier (\cref{fig:JPAnoise}.(a)). Inevitable losses $\eta$ are present between the JPA and the HEMT amplifier. They are mainly caused by the insertion losses of various microwave components such as circulators for example, all thermalised at the same temperature than the JPA ($T_\text{in}$). At a given pump frequency $\omega_\text{p}$, the power spectral density when the JPA is off is given by:
\begin{equation}
PSD_\text{off} = G_\text{HEMT}k_\text{B}( T_\text{HEMT} + T_\text{in}) \simeq k_\text{B} T_\text{HEMT}
\end{equation}
We assume that the input of the JPA is connected to a perfect \SI{50}{\ohm} resistance giving a vacuum noise of half a photon, $T_\text{in}= \hbar \omega_\text{p}/2k_\text{B} =$ \SI{166}{\milli\kelvin}.
We assume as well that the HEMT amplifier is the main source of noise in this case.\\
When the JPA is operated with a gain $G_\text{JPA}$, the $PSD$ increases because of the contribution of the amplified added noise of the JPA, $G_\text{JPA}(1-\eta)k_\text{B} T_\text{JPA}$ with $T_\text{JPA}$ the effective noise temperature of the JPA. The total $PSD$ is then given by:
\begin{equation}
PSD_\text{on} =G_\text{HEMT} k_\text{B} ( T_\text{HEMT}+ G_\text{JPA} (T_\text{JPA}+T_\text{in})(1-\eta) +\eta T_\text{in} )
\end{equation}
where we suppose that the noise of the JPA does not depend on its gain.\\
Therefore, the $PSD$ ratio, $R_{PSD}$ is:
\begin{equation}
	\begin{split}
R_{PSD} =& \frac{PSD_\text{on}}{PSD_\text{off}}\\ =& \frac{T_\text{HEMT}+\eta T_\text{in}}{T_\text{HEMT}+ T_\text{in}} + G_\text{JPA}(1-\eta) \frac{T_\text{JPA}+T_\text{in}}{T_\text{HEMT}+ T_{in}}\\ \simeq& 1 + G_\text{JPA}(1-\eta) \frac{T_\text{JPA}+T_\text{in}}{T_\text{HEMT}}
	\end{split}
\end{equation}
Under these assumptions there is a linear relationship between the gain of the JPA and $R_{PSD}$. This ratio is measured for several gain values (\cref{fig:JPAnoise}.(b)) at frequency $\omega_\text{p}/2\pi=$ \SI{6.913}{\giga\hertz}. From the measured slope ($(1-\eta) (T_\text{JPA}+T_{in})/T_\text{HEMT} \simeq 1/22.7$), we can estimate the noise temperature of the JPA, at frequency $\omega_\text{p}$. Indeed the ratio $T_\text{HEMT}/(1-\eta)$ was measured to be  $\SI{8\pm2}{\kelvin}$ in this setup~\cite{dumur2015}. This translates into $T_\text{JPA}=\SI{180\pm90}{\milli\kelvin}$ corresponding to a number of added photons $ \SI{0.55\pm 0.25}{}$, while the quantum limit stands at $\SI{0.50}{}$~\cite{Caves:1982zz}. Moreover this $PSD$ ratio is comparable to what was reported by other groups~\cite{Frattini:2017ji}.
In conclusion, we present evidence that the JPA reported in this work is near quantum limited but the uncertainty of our measurement prevents us to claim that it performs strictly at the quantum limit.

%\bibliography{Array_JPA}

%

\end{document}